\RequirePackage{ifpdf}
\ifpdf 
\documentclass[pdftex]{sigma}
\else
\documentclass{sigma}
\fi 

\usepackage{upgreek}

\numberwithin{equation}{section}

\newtheorem*{Theorem*}{Theorem}

 { \theoremstyle{definition}

 }

\begin{document}

\renewcommand{\PaperNumber}{***}

\FirstPageHeading

\ShortArticleName{Integrable twelve-component   nonlinear  dynamical  system on a quasi-one-dimensional lattice}

\ArticleName{Integrable twelve-component   nonlinear  dynamical \\ system on a quasi-one-dimensional lattice} 

\Author{Oleksiy O.~VAKHNENKO~$^{\rm a}$ and Vyacheslav O.~VAKHNENKO~$^{\rm b}$}

\AuthorNameForHeading{O.O.~Vakhnenko and V.O.~Vakhnenko}

\Address{$^{\rm a)}$~Department for Theory of Nonlinear Processes in Condensed Matter,
	Bogolyubov  Institute for Theoretical Physics,  The National Academy of Sciences of Ukraine,\\
	14-B Metrologichna Street,  Ky\"{\i}v 03143, Ukraine} 
\EmailD{\href{mailto:email@address}{vakhnenko@bitp.kiev.ua}} 
\URLaddressD{\url{https://orcid.org/0000-0001-8371-9499}}  

\Address{$^{\rm b)}$~Department of Dynamics of Deformable Solids,
	Subbotin Institute of Geophysics,
	The National Academy of Sciences of Ukraine,\\
	63-B Bohdan Khmel$^{,}$nyts$^{,}$kyy Street, Ky\"{\i}v 01054, Ukraine}
\EmailD{\href{mailto:email@address}{vakhnenko@ukr.net}} 
\URLaddressD{\url{https://orcid.org/0000-0002-1250-9563}} 


~\\
{Received 24 May 2025, revised 28 July 2025, accepted 9 September 2025,  Published online 22 October 2025}
~\\

\Abstract{Bearing in mind the potential physical  applicability of multicomponent completely integrable nonlinear dynamical models on quasi-one-dimensional lattices we have developed the novel twelve-component and six-component semi-discrete nonlinear inregrable systems in the framework of semi-discrete Ablowitz--Kaup--Newell--Segur scheme.  The set of lowest local conservation laws found by the generalized direct recurrent technique was shown to be indispensable constructive  tool in the reduction procedure from the prototype to actual field variables.   
Two types of admissible symmetries for the  twelve-component system and one type of symmetry  for the   six-component system have been  established. The mathematical structure  of total local current was shown to support the charge transportation only   by  four of six subsystems incorporated into the  twelve-component system under study. The  twelve-component system is able to model the actions of external parametric drive and external uniform magnetic field via time dependencies and phase factors of coupling parameters.}

\Keywords{Lax integrability; Quasi-one-dimensional  lattice;  Multicomponent system;   Nonlinear dynamics; $\cal{PT}$ symmetry}

\Classification{39A36; 37K10; 35Q55; 58J70} 
 

\tableofcontents

\section{Introduction} 
\label{sec1}
\setcounter{equation}{0} 

Since the fundamental works  by Toda  \cite{toda-JPSJ-22-431,  toda-JPSJ-23-501, toda-PhR-18-1} as well as by Ablowitz and Ladik \cite{ablowitz-JMP-17-1011, ablowitz-SAM-55-213, ablowitz-SAM-58-17} the integrable nonlinear dynamical systems on one-dimensional and quasi-one-dimensional lattices
are steadily gaining a significant influence on modeling a wide variety of  diverse   nonlinear phenomena in  physical, biological and applied sciences. In this respect the multicomponent differential-difference (semi-discrete) nonlinear integrable systems \cite{gerdzhikov-TMP-52-676, tsuchida-JMP-39-4785,  tsuchida-JPA-32-2239, ablowitz-PLA-253-287, vakhnenko-JPA-32-5735, vakhnenko-PRE-60-R2492, ablowitz-CSR-11-159, vakhnenko-PRE-61-7110, vakhnenko-JNMP-24-250, vakhnenko-EPJP-137-1176}   acquire an important place due to their physical importance and mathematical reliability. 

Sometimes, however, the lack of physical imagination  inspires  the semi-discrete nonlinear integrable systems allegedly claimed to be multicomponent but actually composed of several physically uncoupled  one-component or two-component basic systems \cite{carstea-JPA-48-055205, babalic-JPA-50-415201, babalic-RJP-63-114, babalic-IJMPB-34-2050274, yu-WM-94-102500, wen-AML-123-107591, yuan-CSF-168-113180}. The critical overview of semi-discrete nonlinear integrable systems characterized by the false multicomponentness is given in our recent paper \cite{vakhnenko-PLA-452-128460}.

The progress in the development of new differential-differential  multicomponent   nonlinear  integrable  systems  has been prompted by the guiding rules known as the  Ablowitz--Kaup--Newell--Segur scheme \cite{ablowitz-PRL-31-125,  ablowitz-SAM-53-249}. The spatially discretized version of Ablowitz--Kaup--Newell--Segur rules is now widely recognized as the prospective tool for the construction of new differential-difference (semi-discrete) multicomponent   nonlinear  integrable  systems. The key points of spatially discretized  Ablowitz--Kaup--Newell--Segur scheme are summarized in our previous papers \cite{vakhnenko-UJP-69-168,  vakhnenko-UJP-69-577}. We  followed this constructive procedure while developing   the  integrable twelve-component   nonlinear  dynamical system on a quasi-one-dimensional lattice suggested in the present paper.

\section{Semi-discrete zero-curvature equation and the mutually\\ consistent ans\"{a}tze  for the  auxiliary spectral and evolutionary matrices}  
\label{sec2}
\setcounter{equation}{0} 

Any classic   nonlinear  evolutionary system on an infinite quasi-one-dimensional lattice  is said to be integrable in the Lax sense provided it admits  a  matrix-valued  semi-discrete zero-curvature representation \cite{faddeev-HMTS-2007, tu-JPA-23-3903}
\begin{equation}\label{eq-2.1}
	\frac{\mathrm{d}}{\mathrm{d}\tau}L(n|z)=A(n+1|z)L(n|z)-L(n|z)A(n|z)
\end{equation} 
with the auxiliary square matrices $L(n|z)$ and $A(n|z)$ referred to as the spectral and evolutionary matrices, respectively.
Here $\tau$ denotes the continuous  time variable, $n$ stands for the discrete space variable running through the integer numbers from minus infinity to plus infinity, while $z$ marks the time- and space-independent spectral parameter. The spectral matrix $L(n|z)$ is  assumed to be the non-degenerate one ($\det L(n|z)\neq0$).

In this paper we suggest the  auxiliary  matrices $L(n|z)$ and $A(n|z)$ in the following  forms
\begin{eqnarray}\label{eq-2.2}
	L(n|z) = \left(\begin{array}{ccccccccccc}
		0 & t_{12}(n) & u_{13}(n)z^{-1} & 0  \\
		t_{21}(n) & r_{22}(n)z^{2} + t_{22}(n) & s_{23}(n)z + u_{23}(n)z^{-1} & s_{24}(n)z \\
		u_{31}(n)z^{-1} & s_{32}(n)z + u_{32}(n)z^{-1} & t_{33}(n) + v_{33}(n)z^{-2} & t_{34}(n) \\
		0 & s_{42}(n)z & t_{43}(n) & 0 \\	
	\end{array}\right) 
\end{eqnarray} 
\begin{eqnarray}\label{eq-2.3}
	A(n|z) = \left(\begin{array}{ccccccccccc}
		c_{11}(n) & c_{12}(n) & d_{13}(n)z^{-1} & 0  \\
		c_{21}(n) & a_{22}(n)z^{2} + c_{22}(n) & b_{23}(n)z + d_{23}(n)z^{-1} & b_{24}(n)z \\
		d_{31}(n)z^{-1} & b_{32}(n)z + d_{32}(n)z^{-1} & c_{33}(n) + e_{33}(n)z^{-2} & c_{34}(n) \\
		0 & b_{42}(n)z & c_{43}(n) & c_{44}(n) \\
	\end{array}\right) ,
\end{eqnarray} 
where the ansatz (\ref{eq-2.3}) for the evolutionary matrix $A(n|z)$ is extended by the additional terms $c_{11}(n)$ and $c_{44}(n)$ as compared with the ansatz considered  previously \cite{vakhnenko-JNMP-18-401,  vakhnenko-JNMP-18-415}. 

The adopted ans\"{a}tze (\ref{eq-2.2})--(\ref{eq-2.3}) in combination with the zero-curvature equation  (\ref{eq-2.1}) allow to fix  the  majority of  matrix elements $A_{jk}(n|z)$ of evolutionary matrix $A(n|z)$  in terms of prototype field functions $t_{12}(n)$, $u_{13}(n)$, $t_{21}(n)$, $r_{22}(n)$, $t_{22}(n)$, $s_{23}(n)$, $u_{23}(n)$, $s_{24}(n)$, $u_{31}(n)$, $s_{32}(n)$, $u_{32}(n)$, $t_{33}(n)$, $v_{33}(n)$, $t_{34}(n)$, $s_{42}(n)$, $t_{43}(n)$ as well as to  recover the set of primary semi-discrete nonlinear  equations  for the prototype field functions. Under certain plausible assumptions prompted by the on-site local conservation laws the  primary semi-discrete nonlinear  equations are convertible into the one or another  semi-discrete nonlinear   integrable system  of preferable interest. In this sense the suggested ans\"{a}tze (\ref{eq-2.2})--(\ref{eq-2.3}) for  the auxiliary matrices are proved to be mutually consistent. 

N. B. The choice of  proper ansatz  for the evolutionary matrix $A(n|z)$ is not unique and it manifests a particular integrable system in an infinite hierarchy related to the generic spectral matrix  (\ref{eq-2.2}).

\section{Primary semi-discrete nonlinear  equations}  
\label{sec3}
\setcounter{equation}{0}  

To proceed with developing  the semi-discrete nonlinear   integrable systems  of our present interest we have to
rely upon the primary (prototype) set of semi-discrete equations
\begin{eqnarray}\label{eq-3.1}
	\frac{\mathrm{d}}{\mathrm{d}\tau}t_{12}(n)&=&c_{11}(n+1)t_{12}(n)+c_{12}(n+1)t_{22}(n)+d_{13}(n+1)s_{32}(n)\nonumber\\
	&-&t_{12}(n)c_{22}(n)-u_{13}(n)b_{32}(n)
\end{eqnarray}
\begin{eqnarray}\label{eq-3.2}
	\frac{\mathrm{d}}{\mathrm{d}\tau}u_{13}(n)&=&c_{11}(n+1)u_{13}(n)+c_{12}(n+1)u_{23}(n)+d_{13}(n+1)t_{33}(n)\nonumber\\
	&-&t_{12}(n)d_{23}(n)-u_{13}(n)c_{33}(n)
\end{eqnarray}
\begin{eqnarray}\label{eq-3.3}
	\frac{\mathrm{d}}{\mathrm{d}\tau}t_{21}(n)&=&c_{22}(n+1)t_{21}(n)+b_{23}(n+1)u_{31}(n)\nonumber\\
	&-&t_{21}(n)c_{11}(n)-t_{22}(n)c_{21}(n)-s_{23}(n)d_{31}(n)
\end{eqnarray}
\begin{eqnarray}\label{eq-3.4}
	\frac{\mathrm{d}}{\mathrm{d}\tau}r_{22}(n)&=&c_{22}(n+1)r_{22}(n)+b_{23}(n+1)s_{32}(n)+b_{24}(n+1)s_{42}(n) \nonumber\\
&-&r_{22}(n)c_{22}(n)-s_{23}(n)b_{32}(n)-s_{24}(n)b_{42}(n)
\end{eqnarray}
\begin{eqnarray}\label{eq-3.5}
	\frac{\mathrm{d}}{\mathrm{d}\tau}t_{22}(n)&=&c_{21}(n+1)t_{12}(n)+c_{22}(n+1)t_{22}(n)-t_{21}(n)c_{12}(n)-t_{22}(n)c_{22}(n) \nonumber\\
	&+&b_{23}(n+1)u_{32}(n)+d_{23}(n+1)s_{32}(n)-s_{23}(n)d_{32}(n)-u_{23}(n)b_{32}(n)
\end{eqnarray}
\begin{eqnarray}\label{eq-3.6}
	\frac{\mathrm{d}}{\mathrm{d}\tau}s_{23}(n)&=&a_{22}(n+1)u_{23}(n)+c_{22}(n+1)s_{23}(n)-r_{22}(n)d_{23}(n)-t_{22}(n)b_{23}(n) \nonumber\\
	&+&b_{23}(n+1)t_{33}(n)+b_{24}(n+1)t_{43}(n)-s_{23}(n)c_{33}(n)-s_{24}(n)c_{43}(n)
\end{eqnarray}
\begin{eqnarray}\label{eq-3.7}
	\frac{\mathrm{d}}{\mathrm{d}\tau}u_{23}(n)&=&c_{21}(n+1)u_{13}(n)+c_{22}(n+1)u_{23}(n)-t_{21}(n)d_{13}(n)-t_{22}(n)d_{23}(n) \nonumber\\
	&+&b_{23}(n+1)v_{33}(n)+d_{23}(n+1)t_{33}(n)-s_{23}(n)e_{33}(n)-u_{23}(n)c_{33}(n)
\end{eqnarray}
\begin{eqnarray}\label{eq-3.8}
	\frac{\mathrm{d}}{\mathrm{d}\tau}s_{24}(n)&=&c_{22}(n+1)s_{24}(n)+b_{23}(n+1)t_{34}(n)\nonumber\\
	&-&t_{22}(n)b_{24}(n)-s_{23}(n)c_{34}(n)-s_{24}(n)c_{44}(n)
\end{eqnarray}
\begin{eqnarray}\label{eq-3.9}
	\frac{\mathrm{d}}{\mathrm{d}\tau}u_{31}(n)&=&d_{32}(n+1)t_{21}(n)+c_{33}(n+1)u_{31}(n)\nonumber\\
	&-&u_{31}(n)c_{11}(n)-u_{32}(n)c_{21}(n)-t_{33}(n)d_{31}(n)
\end{eqnarray}
\begin{eqnarray}\label{eq-3.10}
	\frac{\mathrm{d}}{\mathrm{d}\tau}s_{32}(n)&=&b_{32}(n+1)t_{22}(n)+d_{32}(n+1)r_{22}(n)-s_{32}(n)c_{22}(n)-u_{32}(n)a_{22}(n) \nonumber\\
	&+&c_{33}(n+1)s_{32}(n)+c_{34}(n+1)s_{42}(n)-t_{33}(n)b_{32}(n)-t_{34}(n)b_{42 }(n)
\end{eqnarray}
\begin{eqnarray}\label{eq-3.11}
	\frac{\mathrm{d}}{\mathrm{d}\tau}u_{32}(n)&=&d_{31}(n+1)t_{12}(n)+d_{32}(n+1)t_{22}(n)-u_{31}(n)c_{12}(n)-u_{32}(n)c_{22}(n) \nonumber\\
	&+&c_{33}(n+1)u_{32}(n)+e_{33}(n+1)s_{32}(n)-t_{33}(n)d_{32}(n)-v_{33}(n)b_{32}(n)
\end{eqnarray}
\begin{eqnarray}\label{eq-3.12}
	\frac{\mathrm{d}}{\mathrm{d}\tau}t_{33}(n)&=&b_{32}(n+1)u_{23}(n)+d_{32}(n+1)s_{23}(n)-s_{32}(n)d_{23}(n)-u_{32}(n)b_{23}(n) \nonumber\\
	&+&c_{33}(n+1)t_{33}(n)+c_{34}(n+1)t_{43}(n)-t_{33}(n)c_{33}(n)-t_{34}(n)c_{43}(n)
\end{eqnarray}
\begin{eqnarray}\label{eq-3.13}
	\frac{\mathrm{d}}{\mathrm{d}\tau}v_{33}(n)&=&d_{31}(n+1)u_{13}(n)+d_{32}(n+1)u_{23}(n)+c_{33}(n+1)v_{33}(n) \nonumber\\
	&-&u_{31}(n1)d_{13}(n)-u_{32}(n)d_{23}(n)-v_{33}(n)c_{33}(n)
\end{eqnarray}
\begin{eqnarray}\label{eq-3.14}
	\frac{\mathrm{d}}{\mathrm{d}\tau}t_{34}(n)&=&d_{32}(n+1)s_{24}(n)+c_{33}(n+1)t_{34}(n)\nonumber\\ 
	&-&u_{32}(n)b_{24}(n)-t_{33}(n)c_{34}(n)-t_{34}(n)c_{44}(n)
\end{eqnarray}
\begin{eqnarray}\label{eq-3.15}
	\frac{\mathrm{d}}{\mathrm{d}\tau}s_{42}(n)&=&b_{42}(n+1)t_{22}(n)+c_{43}(n+1)s_{32}(n)+c_{44}(n+1)s_{42}(n)\nonumber\\
	&-&s_{42}(n)c_{22}(n)-t_{43}(n)b_{32}(n)
\end{eqnarray}
\begin{eqnarray}\label{eq-3.16}
	\frac{\mathrm{d}}{\mathrm{d}\tau}t_{43}(n)&=&b_{42}(n+1)u_{23}(n)+c_{43}(n+1)t_{33}(n)+c_{44}(n+1)t_{43}(n)\nonumber\\
	&-&s_{42}(n)d_{23}(n)-t_{43}(n)c_{33}(n)\,.
\end{eqnarray}
Here fourteen functions specifying the matrix elements $A_{jk}(n|z)$ are fixed by formulas 
\begin{eqnarray}\label{eq-3.17}
	a_{22}(n)=a_{22}
\end{eqnarray}
\begin{eqnarray}\label{eq-3.18}
	c_{21}(n)=a_{22}t_{21}(n)/r_{22}(n)
\end{eqnarray}
\begin{eqnarray}\label{eq-3.19}
	c_{12}(n+1)=t_{12}(n)a_{22}/r_{22}(n)
\end{eqnarray}
\begin{eqnarray}\label{eq-3.20}
	b_{23}(n)=a_{22}s_{23}(n)/r_{22}(n)
\end{eqnarray}
\begin{eqnarray}\label{eq-3.21}
	b_{32}(n+1)=s_{32}(n)a_{22}/r_{22}(n)
\end{eqnarray}
\begin{eqnarray}\label{eq-3.22}
	b_{24}(n)=a_{22}s_{24}(n)/r_{22}(n)
\end{eqnarray}
\begin{eqnarray}\label{eq-3.23}
	b_{42}(n+1)=s_{42}(n)a_{22}/r_{22}(n)
\end{eqnarray}
\begin{eqnarray}\label{eq-3.24}
	e_{33}(n)=e_{33}
\end{eqnarray}
\begin{eqnarray}\label{eq-3.25}
	c_{34}(n)=e_{33}t_{34}(n)/v_{33}(n)
\end{eqnarray}
\begin{eqnarray}\label{eq-3.26}
	c_{43}(n+1)=t_{43}(n)e_{33}/v_{33}(n)
\end{eqnarray}
\begin{eqnarray}\label{eq-3.27}
	d_{32}(n)=e_{33}u_{32}(n)/v_{33}(n)
\end{eqnarray}
\begin{eqnarray}\label{eq-3.28}
	d_{23}(n+1)=u_{23}(n)e_{33}/v_{33}(n)
\end{eqnarray}
\begin{eqnarray}\label{eq-3.29}
	d_{31}(n)=e_{33}u_{31}(n)/v_{33}(n)
\end{eqnarray}
\begin{eqnarray}\label{eq-3.30}
	d_{13}(n+1)=u_{13}(n)e_{33}/v_{33}(n)\,,
\end{eqnarray}
whilst the rest four functions $c_{11}(n)$, $c_{22}(n)$, $c_{33}(n)$, $c_{44}(n)$ remain as yet being unfixed. 

The above written sixteen semi-discrete equations (\ref{eq-3.1})--(\ref{eq-3.16}) and fourteen specification formulas (\ref{eq-3.17})--(\ref{eq-3.30}) are obtainable by the simple algebraic manipulations involving the zero-curvature equation  (\ref{eq-2.1}) and the ans\"{a}tze (\ref{eq-2.2})--(\ref{eq-2.3}) for the auxiliary matrices.  

In general the spatially independent parameters $a_{22}$ and $e_{33}$ can  be arbitrary functions of time. Thus, the obtained set of semi-discrete nonlinear  equations (\ref{eq-3.1})--(\ref{eq-3.16}) deciphered by the  shorthand  formulas (\ref{eq-3.17})--(\ref{eq-3.30}) can in general be the parametrically driven one. Such an evident but physically important  property potentially  admissible  for a wide class of semi-discrete nonlinear integrable systems  is usually overlooked or ignored by the scientific community.

\section{Constructive part of the on-site  local conservation laws}  
\label{sec4}
\setcounter{equation}{0} 

The most constructive way to  fix four  arbitrary  functions $c_{11}(n)$, $c_{22}(n)$, $c_{33}(n)$, $c_{44}(n)$ is based on the use of lowest local conservation laws 
\begin{eqnarray}\label{eq-4.1}
	\frac{\mathrm{d}}{\mathrm{d}\tau}\rho_{11}(n)=c_{11}(n+1)+c_{22}(n+1)+c_{33}(n+1) -c_{11}(n)-c_{22}(n)-c_{33}(n)
\end{eqnarray}
\begin{eqnarray}\label{eq-4.2}
	\frac{\mathrm{d}}{\mathrm{d}\tau}\rho_{22}(n)&=&c_{22}(n+1) + a_{22}\frac{s_{23}(n+1)s_{32}(n)}{r_{22}(n+1)r_{22}(n)} + a_{22}\frac{s_{24}(n+1)s_{42}(n)}{r_{22}(n+1)r_{22}(n)}\nonumber\\ &-&c_{22}(n)-a_{22}\frac{s_{23}(n)s_{32}(n-1)}{r_{22}(n)r_{22}(n-1)}- a_{22}\frac{s_{24}(n)s_{42}(n-1)}{r_{22}(n)r_{22}(n-1)}
\end{eqnarray}
\begin{eqnarray}\label{eq-4.3}
	\frac{\mathrm{d}}{\mathrm{d}\tau}\rho(n)&=&c_{11}(n+1)+c_{22}(n+1)+c_{33}(n+1)+c_{44}(n+1)\nonumber\\ &-&c_{11}(n)-c_{22}(n)-c_{33}(n)-c_{44}(n)
\end{eqnarray}
\begin{eqnarray}\label{eq-4.4}
	\frac{\mathrm{d}}{\mathrm{d}\tau}\rho_{33}(n)&=&c_{33}(n+1) + e_{33}\frac{u_{32}(n+1)u_{23}(n)}{v_{33}(n+1)v_{33}(n)} + e_{33}\frac{u_{31}(n+1)u_{13}(n)}{v_{33}(n+1)v_{33}(n)}\nonumber\\ &-&c_{33}(n)-e_{33}\frac{u_{32}(n)u_{23}(n-1)}{v_{33}(n)v_{33}(n-1)}- e_{33}\frac{u_{31}(n)u_{13}(n-1)}{v_{33}(n)v_{33}(n-1)}
\end{eqnarray}
\begin{eqnarray}\label{eq-4.5}
	\frac{\mathrm{d}}{\mathrm{d}\tau}\rho_{44}(n)=c_{22}(n+1)+c_{33}(n+1)+c_{44}(n+1) -c_{22}(n)-c_{33}(n)-c_{44}(n)
\end{eqnarray}
associated with the respective on-site local conserved  densities 
\begin{eqnarray}\label{eq-4.6}
   \rho_{11}(n)=\ln[t_{12}(n)v_{33}(n)t_{21}(n)&+&u_{13}(n)t_{22}(n)u_{31}(n)\nonumber\\
	&-&t_{12}(n)u_{23}(n)u_{31}(n)-u_{13}(n)u_{32}(n)t_{21}(n)]
\end{eqnarray}
\begin{eqnarray}\label{eq-4.7}
	\rho_{22}(n)=\ln[r_{22}(n)]
\end{eqnarray} 
\begin{eqnarray}\label{eq-4.8}
	\rho(n)=\ln[u_{13}(n)s_{42}(n)-t_{12}(n)t_{43}(n)]+\ln[u_{31}(n)s_{24}(n)-t_{21}(n)t_{34}(n)]
\end{eqnarray} 
\begin{eqnarray}\label{eq-4.9}
	\rho_{33}(n)=\ln[v_{33}(n)]
\end{eqnarray} 
\begin{eqnarray}\label{eq-4.10}
	\rho_{44}(n)=\ln[t_{43}(n)r_{22}(n)t_{34}(n)&+&s_{42}(n)t_{33}(n)s_{24}(n)\nonumber\\
	&-&t_{43}(n)s_{32}(n)s_{24}(n)-s_{42}(n)s_{23}(n)t_{34}(n)]\,.
\end{eqnarray}
These ten formulas (\ref{eq-4.1})--(\ref{eq-4.10}) are obtainable mainly within the  modified direct recursive procedure \cite{vakhnenko-JNMP-18-401, vakhnenko-JNMP-18-415, vakhnenko-JNMP-20-606, vakhnenko-JMP-56-033505} originated  as the  generalization of Tsuchida--Ujino--Wadati  direct recursive approach \cite{tsuchida-JMP-39-4785, tsuchida-JPA-32-2239}. Only two of them, namely formulas (\ref{eq-4.3}) and (\ref{eq-4.8}), arise as   a simple  paraphrase  of so-called universal local conservation law 
\begin{equation}\label{eq-4.11}
	\frac{\mathrm{d}}{\mathrm{d}\tau}\ln\left[\det L(n|z)\right]=\mathrm{Sp} A(n+1|z) -\mathrm{Sp} A(n|z)
\end{equation} 
following directly from the zero-curvature equation (\ref{eq-2.1}). 

In order to fix  four arbitrary functions $c_{11}(n)$, $c_{22}(n)$, $c_{33}(n)$, $c_{44}(n)$
we must impose four  constraints onto the left-hand sides of five on-site local conservation laws (\ref{eq-4.1})--(\ref{eq-4.5}). Of course, such a procedure admits  a number of  diverse realizations giving rise to one or another  particular sample of semi-discrete nonlinear integrable system. For this reason, the  functions $c_{11}(n)$, $c_{22}(n)$, $c_{33}(n)$, $c_{44}(n)$ can be  referred to as the sampling ones.

Meanwhile,  our previous papers \cite{vakhnenko-JNMP-18-401,  vakhnenko-JNMP-18-415} have substantially  restricted the range of claimed  diversity by ignoring the functions $c_{11}(n)$ and $c_{44}(n)$ in  ansatz for the evolutionary matrix  $A(n|z)$. In our present consideration this ignorance is tantamount to the following two  constrains 
\begin{eqnarray}\label{eq-4.12}
	\frac{\mathrm{d}}{\mathrm{d}\tau}\rho_{11}(n)=\frac{\mathrm{d}}{\mathrm{d}\tau}\rho(n)=\frac{\mathrm{d}}{\mathrm{d}\tau}\rho_{44}(n)\,.
\end{eqnarray}
The variability of another two admissible constrains has been  thoroughly analyzed in the second  \cite{vakhnenko-JNMP-18-415} of  just mentioned   papers. 

In what follows we try to grasp the key features  of a particular  semi-discrete nonlinear integrable system specified   by  four the most natural   demands 
\begin{eqnarray}\label{eq-4.13}
	\frac{\mathrm{d}}{\mathrm{d}\tau}\rho_{22}(n)=0=\frac{\mathrm{d}}{\mathrm{d}\tau}\rho_{33}(n)
\end{eqnarray}
\begin{eqnarray}\label{eq-4.14}
	\frac{\mathrm{d}}{\mathrm{d}\tau}\rho_{11}(n)=0=\frac{\mathrm{d}}{\mathrm{d}\tau}\rho_{44}(n)\,.
\end{eqnarray}
Then the local conservation laws  (\ref{eq-4.1})--(\ref{eq-4.2}) and (\ref{eq-4.4})--(\ref{eq-4.5}) yield 
\begin{eqnarray}\label{eq-4.15}
	c_{11}(n)=c_{11}&+&a_{22}\frac{s_{23}(n)s_{32}(n-1)}{r_{22}(n)r_{22}(n-1)}+ a_{22}\frac{s_{24}(n)s_{42}(n-1)}{r_{22}(n)r_{22}(n-1)}\nonumber\\ &+&e_{33}\frac{u_{32}(n)u_{23}(n-1)}{v_{33}(n)v_{33}(n-1)}+ e_{33}\frac{u_{31}(n)u_{13}(n-1)}{v_{33}(n)v_{33}(n-1)}
\end{eqnarray}
\begin{eqnarray}\label{eq-4.16}
	c_{22}(n)=c_{22}-a_{22}\frac{s_{23}(n)s_{32}(n-1)}{r_{22}(n)r_{22}(n-1)}- a_{22}\frac{s_{24}(n)s_{42}(n-1)}{r_{22}(n)r_{22}(n-1)}
\end{eqnarray}
\begin{eqnarray}\label{eq-4.17}
    c_{33}(n)=c_{33}-e_{33}\frac{u_{32}(n)u_{23}(n-1)}{v_{33}(n)v_{33}(n-1)}- e_{33}\frac{u_{31}(n)u_{13}(n-1)}{v_{33}(n)v_{33}(n-1)} 
\end{eqnarray}
\begin{eqnarray}\label{eq-4.18}
	c_{44}(n)=c_{44}&+&e_{33}\frac{u_{32}(n)u_{23}(n-1)}{v_{33}(n)v_{33}(n-1)}+ e_{33}\frac{u_{31}(n)u_{13}(n-1)}{v_{33}(n)v_{33}(n-1)}\nonumber\\
	&+&a_{22}\frac{s_{23}(n)s_{32}(n-1)}{r_{22}(n)r_{22}(n-1)}+ a_{22}\frac{s_{24}(n)s_{42}(n-1)}{r_{22}(n)r_{22}(n-1)}\,.
\end{eqnarray} 
Here the time-dependent  free parameters $c_{11}$, $c_{22}$, $c_{33}$, $c_{44}$ can be safely  removed by the proper gauge transformations of field functions. Therefore, without the loss of generality we equalize each of   four parameters $c_{jj}$   to  zero.

As a result of adopted constraints (\ref{eq-4.13}) and (\ref{eq-4.14}) the number of independent field functions is reduced from the sixteen to the twelve ones. The choice of  functions  $t_{12}(n)$, $u_{13}(n)$, $t_{21}(n)$,   $s_{23}(n)$, $u_{23}(n)$, $s_{24}(n)$, $u_{31}(n)$, $s_{32}(n)$, $u_{32}(n)$,  $t_{34}(n)$, $s_{42}(n)$, $t_{43}(n)$  as being truly independent appears to be the most convenient. Thus, the functions $r_{22}(n)$, $t_{22}(n)$ and $v_{33}(n)$, $t_{33}(n)$ acquire the status of concomitant functions. 

The arbitrary spatial dependencies of time independent concomitant functions $r_{22}(n)$ and $v_{33}(n)$ could in principle imitate the action of external substrate, but  here  we discard  this idea in order to preserve the uniformity of space. Thus, the concomitant functions $r_{22}(n)$ and $v_{33}(n)$ are reduced to the sheer constant parameters. The subsequent  proper scaling procedure of field functions and the coupling parameters  $a_{22}$ and $e_{33}$ gives rise to the equations of motion invariant to the primary equations of motion specified by  the equalities 
\begin{eqnarray}\label{eq-4.19}
	r_{22}(n)=1=v_{33}(n)\,. 
\end{eqnarray}

As to the  concomitant functions $t_{22}(n)$ and $t_{33}(n)$, they are determined by the simple algebraic   equations  
\begin{eqnarray}\label{eq-4.20}
   t_{12}(n)v_{33}(n)t_{21}(n)&+&u_{13}(n)t_{22}(n)u_{31}(n)\nonumber\\
    &-&t_{12}(n)u_{23}(n)u_{31}(n)-u_{13}(n)u_{32}(n)t_{21}(n) = T_{22}(n)
\end{eqnarray} 
and 
\begin{eqnarray}\label{eq-4.21}
	t_{43}(n)r_{22}(n)t_{34}(n)&+&s_{42}(n)t_{33}(n)s_{24}(n)\nonumber\\
	&-&t_{43}(n)s_{32}(n)s_{24}(n)-s_{42}(n)s_{23}(n)t_{34}(n) = T_{33}(n)  
\end{eqnarray}
in accordance  with the adopted differential constraints (\ref{eq-4.14}) accompanied by the  explicit expressions  (\ref{eq-4.6}) and (\ref{eq-4.10}) for the on-site local conserved densities $\rho_{11}(n)$ and $\rho_{44}(n)$. 
Here $T_{22}(n)$ and $T_{33}(n)$ are time-independent arbitrary functions of spatial variable $n$. 

In the present research we restrict ourselves  to the simplest possible variant 
\begin{eqnarray}\label{eq-4.22}
	T_{22}(n) = 0 = T_{33}(n) 
\end{eqnarray}
supplemented by the earlier adopted normalizations (\ref{eq-4.19}) for $r_{22}(n)$ and $v_{33}(n)$. In addition, we  introduce the transformation formulas
\begin{eqnarray}\label{eq-4.23}
	t_{12}(n)=u_{13}(n)\bar{u}_{32}(n)
\end{eqnarray} 
\begin{eqnarray}\label{eq-4.24}
	t_{21}(n)=\bar{u}_{23}(n)u_{31}(n)
\end{eqnarray}
\begin{eqnarray}\label{eq-4.25}
	t_{43}(n)=s_{42}(n)\bar{s}_{23}(n)
\end{eqnarray}
\begin{eqnarray}\label{eq-4.26}
	t_{34}(n)=\bar{s}_{32}(n)s_{24}(n)
\end{eqnarray}
serving to replace  the field functions  $t_{12}(n)$,  $t_{21}(n)$, $t_{43}(n)$, $t_{34}(n)$  by the  more suitable ones  $\bar{u}_{32}(n)$,  $\bar{u}_{23}(n)$, $\bar{s}_{23}(n)$, $\bar{s}_{32}(n)$. Then, the  equations  (\ref{eq-4.20}) and (\ref{eq-4.21}) for the concomitant functions $t_{22}(n)$  and   $t_{33}(n)$  yield very simple  results
\begin{eqnarray}\label{eq-4.27}
	t_{22}(n)=u_{23}(n)\bar{u}_{32}(n)+\bar{u}_{23}(n)u_{32}(n)-\bar{u}_{23}(n)\bar{u}_{32}(n)
\end{eqnarray} 
\begin{eqnarray}\label{eq-4.28}
	t_{33}(n)=s_{23}(n)\bar{s}_{32}(n)\,+\,\bar{s}_{23}(n)s_{32}(n)\,-\,\bar{s}_{23}(n)\bar{s}_{32}(n)\,.
\end{eqnarray}

We finalize this Section by presenting four  important local conserved densities  \cite{vakhnenko-JNMP-18-401,  vakhnenko-JNMP-18-415}
\begin{equation}\label{eq-4.29}
	\rho^{+}_{22}(n)=	\frac{t_{22}(n)}{r_{22}(n)} + \frac{s_{23}(n+1)s_{32}(n)}{r_{22}(n+1)r_{22}(n)} + \frac{s_{24}(n+1)s_{42}(n)}{r_{22}(n+1)r_{22}(n)}
\end{equation} 
\begin{equation}\label{eq-4.30}
	\rho^{-}_{22}(n)=	\frac{t_{22}(n)}{r_{22}(n)} + \frac{s_{23}(n)s_{32}(n-1)}{r_{22}(n)r_{22}(n-1)} + \frac{s_{24}(n)s_{42}(n-1)}{r_{22}(n)r_{22}(n-1)}
\end{equation} 
\begin{equation}\label{eq-4.31}
	\rho^{+}_{33}(n)=	\frac{t_{33}(n)}{v_{33}(n)} + \frac{u_{32}(n+1)u_{23}(n)}{v_{33}(n+1)v_{33}(n)} + \frac{u_{31}(n+1)u_{13}(n)}{v_{33}(n+1)v_{33}(n)}
\end{equation} 
\begin{equation}\label{eq-4.32}
	\rho^{-}_{33}(n)=	\frac{t_{33}(n)}{v_{33}(n)} + \frac{u_{32}(n)u_{23}(n-1)}{v_{33}(n)v_{33}(n-1)} + \frac{u_{31}(n)u_{13}(n-1)}{v_{33}(n)v_{33}(n-1)}
\end{equation} 
that can be useful in constructing the appropriate Hamiltonian function for one or another particular realization of semi-discrete nonlinear integrable system  associated with the  adopted ans\"{a}tze (\ref{eq-2.2})--(\ref{eq-2.3}) for the auxiliary spectral and evolutionary matrices.

The above listed local conserved densities  (\ref{eq-4.29})--(\ref{eq-4.32}) are written in the most general form admitting any feasible fixation of sampling functions $c_{jj}(n)$.

\section{Appropriate  reductions of field functions and coupling\\ parameters}  
\label{sec5}
\setcounter{equation}{0}

Now   all preliminary preparations have been  completed  and we are able  to make  appropriate reductions  in   the prototype set  of semi-discrete equations (\ref{eq-3.1})--(\ref{eq-3.16}) by having taken into account the specification formulas (\ref{eq-3.17})--(\ref{eq-3.30}),  (\ref{eq-4.15})--(\ref{eq-4.18})   for the  functional  elements of evolutionary matrix,  the transformation formulas (\ref{eq-4.23})--(\ref{eq-4.26}) for more suitable field functions, as well as the expressions  (\ref{eq-4.19})--(\ref{eq-4.22}) for the concomitant quantities. 

The proper consideration gives rise to the two types of reductions characterized by the twelve and six actual field functions, respectively. 
In both of the announced  reduction procedures the parameter $\sigma$ is set to distinguish two admissible types of nonlinearities, namely,  the  attractive (focusing)  $\sigma=+1$ and repulsive (defocusing)  $\sigma=-1$ ones.

\subsection{Reduction to   twelve   field functions and two coupling parameters}
The first type of reductions is specified by the following formulas
\begin{eqnarray}\label{eq-5.1}
	s_{23}(n)=+q_{+}(n)\qquad\qquad\qquad\qquad\qquad u_{32}(n)=-\sigma r_{+}(n)
\end{eqnarray}
\begin{eqnarray}\label{eq-5.2}
	u_{23}(n)=-q_{-}(n)\qquad\qquad\qquad\qquad\qquad s_{32}(n)=+\sigma r_{-}(n)
\end{eqnarray}
\begin{eqnarray}\label{eq-5.3}
	\bar{s}_{23}(n)=+\bar{q}_{+}(n)\qquad\qquad\qquad\qquad\qquad \bar{u}_{32}(n)=-\sigma \bar{r}_{+}(n)
\end{eqnarray}
\begin{eqnarray}\label{eq-5.4}
	\bar{u}_{23}(n)=-\bar{q}_{-}(n)\qquad\qquad\qquad\qquad\qquad \bar{s}_{32}(n)=+\sigma \bar{r}_{-}(n)
\end{eqnarray}
\begin{eqnarray}\label{eq-5.5}
	s_{24}(n)=+f_{+}(n)\qquad\qquad\qquad\qquad\qquad u_{31}(n)=-\sigma g_{+}(n)
\end{eqnarray}
\begin{eqnarray}\label{eq-5.6}
	u_{13}(n)=-f_{-}(n)\qquad\qquad\qquad\qquad\qquad s_{42}(n)=+\sigma g_{-}(n) 
\end{eqnarray}
\begin{eqnarray}\label{eq-5.7}
	t_{22}(n)=+\sigma \mu(n)\qquad\qquad\qquad\qquad\qquad\, t_{33}(n)=+\sigma \nu(n)
\end{eqnarray}
\begin{eqnarray}\label{eq-5.8}
	a_{22}=-\mathrm{i}\alpha\qquad\qquad\qquad\qquad\quad\qquad\qquad\, e_{33}=+\mathrm{i}\beta \,.  
\end{eqnarray}

The twelve-component semi-discrete nonlinear integrable system written in terms of above introduced quantities (\ref{eq-5.1})--(\ref{eq-5.8}) is presented in Section 6.

\subsection{Reduction to  six field functions and  one coupling parameter}
The second type of reductions is specified by the following formulas
\begin{eqnarray}\label{eq-5.9}
s_{23}(n)=+w_{+}(n)\qquad\qquad\qquad u_{32}(n)=-\sigma w_{+}(n)
\end{eqnarray}
\begin{eqnarray}\label{eq-5.10}
u_{23}(n)=-w_{-}(n)\qquad\qquad\qquad s_{32}(n)=+\sigma w_{-}(n)
\end{eqnarray}
\begin{eqnarray}\label{eq-5.11}
\bar{s}_{23}(n)=+\bar{w}_{+}(n)\qquad\qquad\qquad \bar{u}_{32}(n)=-\sigma \bar{w}_{+}(n)
\end{eqnarray}
\begin{eqnarray}\label{eq-5.12}
\bar{u}_{23}(n)=-\bar{w}_{-}(n)\qquad\qquad\qquad \bar{s}_{32}(n)=+\sigma \bar{w}_{-}(n)
\end{eqnarray}
\begin{eqnarray}\label{eq-5.13}
s_{24}(n)=+h_{+}(n)\qquad\qquad\qquad u_{31}(n)=-\sigma h_{+}(n)
\end{eqnarray}
\begin{eqnarray}\label{eq-5.14}
u_{13}(n)=-h_{-}(n)\qquad\qquad\qquad s_{42}(n)=+\sigma h_{-}(n) 
\end{eqnarray}
\begin{eqnarray}\label{eq-5.15}
t_{22}(n)=+\sigma \pi(n)\qquad\qquad\qquad~  t_{33}(n)=+\sigma \pi(n) 
\end{eqnarray}
\begin{eqnarray}\label{eq-5.16}
a_{22}=\varkappa\qquad\qquad\qquad\qquad\qquad\, e_{33}(n)=\varkappa \,.
\end{eqnarray}

The six-component semi-discrete nonlinear integrable system written in terms of above introduced quantities (\ref{eq-5.9})--(\ref{eq-5.16}) is presented in Section 7.\\

\section{Twelve-component semi-discrete  nonlinear  integrable  system and its admissible symmetries}  
\label{sec6}
\setcounter{equation}{0}

The reduction formulas (\ref{eq-5.1})--(\ref{eq-5.8}) listed in {\it Subsection 5.1} as applied to the prototype set  of semi-discrete equations (\ref{eq-3.1})--(\ref{eq-3.16})
accompanied by the formulas referred in the first paragraph of Section 5 give rise to the following twelve-component semi-discrete  nonlinear  integrable  system 
\begin{eqnarray}\label{eq-6.1}
	 && +\mathrm{i}\frac{\mathrm{d}}{\mathrm{d}\tau}q_{+}(n) = -\alpha q_{-}(n) - \beta  q_{-}(n-1)[1+\sigma r_{+}(n)q_{+}(n)] - \alpha\sigma\mu(n)q_{+}(n)\nonumber\\ 
	 && +\alpha\sigma q_{+}(n+1)[\nu(n)-r_{-}(n)q_{+}(n)] - \alpha\sigma f_{+}(n+1)g_{-}(n)[q_{+}(n)-\bar{q}_{+}(n)]\nonumber\\
	 && -\beta\sigma f_{-}(n-1)g_{+}(n)q_{+}(n) +\beta\sigma f_{+}(n)g_{-}(n-1)\bar{q}_{+}(n-1)
\end{eqnarray}
\begin{eqnarray}\label{eq-6.2}
	&& -\mathrm{i}\frac{\mathrm{d}}{\mathrm{d}\tau}r_{+}(n) = -\beta r_{-}(n) - \alpha  r_{-}(n-1)[1+\sigma q_{+}(n)r_{+}(n)] - \beta\sigma\nu(n)r_{+}(n)\nonumber\\ 
	&& +\beta\sigma r_{+}(n+1)[\mu(n)-q_{-}(n)r_{+}(n)] - \beta\sigma g_{+}(n+1)f_{-}(n)[r_{+}(n)-\bar{r}_{+}(n)]\nonumber\\
	&& -\alpha\sigma g_{-}(n-1)f_{+}(n)r_{+}(n) +\alpha\sigma g_{+}(n)f_{-}(n-1)\bar{r}_{+}(n-1)
\end{eqnarray}
\begin{eqnarray}\label{eq-6.3}
	&& +\mathrm{i}\frac{\mathrm{d}}{\mathrm{d}\tau}q_{-}(n) = -\beta q_{+}(n) - \alpha  q_{+}(n+1)[1+\sigma r_{-}(n)q_{-}(n)] - \beta\sigma\nu(n)q_{-}(n)\nonumber\\ 
	&& +\beta\sigma q_{-}(n-1)[\mu(n)-r_{+}(n)q_{-}(n)] - \beta\sigma f_{-}(n-1)g_{+}(n)[q_{-}(n)-\bar{q}_{-}(n)]\nonumber\\
	&& -\alpha\sigma f_{+}(n+1)g_{-}(n)q_{-}(n) +\alpha\sigma f_{-}(n)g_{+}(n+1)\bar{q}_{-}(n+1)
\end{eqnarray}
\begin{eqnarray}\label{eq-6.4}
	&& -\mathrm{i}\frac{\mathrm{d}}{\mathrm{d}\tau}r_{-}(n) = -\alpha r_{+}(n) - \beta  r_{+}(n+1)[1+\sigma q_{-}(n)r_{-}(n)] - \alpha\sigma\mu(n)r_{-}(n)\nonumber\\ 
	&& +\alpha\sigma r_{-}(n-1)[\nu(n)-q_{+}(n)r_{-}(n)] - \alpha\sigma g_{-}(n-1)f_{+}(n)[r_{-}(n)-\bar{r}_{-}(n)]\nonumber\\
	&& -\beta\sigma g_{+}(n+1)f_{-}(n)r_{-}(n) +\beta\sigma g_{-}(n)f_{+}(n+1)\bar{r}_{-}(n+1)
\end{eqnarray}    
	
\begin{eqnarray}\label{eq-6.5}
	&& +\mathrm{i}\frac{\mathrm{d}}{\mathrm{d}\tau}\bar{q}_{+}(n) = -\alpha q_{-}(n) - \beta  q_{-}(n-1)[1+\sigma r_{+}(n)\bar{q}_{+}(n)] - \alpha\sigma\mu(n)\bar{q}_{+}(n)\nonumber\\ 
	&& -\beta\sigma \bar{q}_{+}(n)[\nu(n)-r_{-}(n)\bar{q}_{+}(n)] - \alpha\sigma \bar{q}_{+}(n)r_{-}(n-1)[q_{+}(n)-\bar{q}_{+}(n)]\nonumber\\
	&& -\alpha\sigma f_{+}(n)g_{-}(n-1)\bar{q}_{+}(n) -\beta\sigma f_{-}(n-1)g_{+}(n)\bar{q}_{+}(n)
\end{eqnarray}
\begin{eqnarray}\label{eq-6.6}
	&& -\mathrm{i}\frac{\mathrm{d}}{\mathrm{d}\tau}\bar{r}_{+}(n) = -\beta r_{-}(n) - \alpha r_{-}(n-1)[1+\sigma q_{+}(n)\bar{r}_{+}(n)] - \beta\sigma\nu(n)\bar{r}_{+}(n)\nonumber\\ 
	&& -\alpha\sigma \bar{r}_{+}(n)[\mu(n)-q_{-}(n)\bar{r}_{+}(n)] - \beta\sigma \bar{r}_{+}(n)q_{-}(n-1)[r_{+}(n)-\bar{r}_{+}(n)]\nonumber\\
	&& -\beta\sigma g_{+}(n)f_{-}(n-1)\bar{r}_{+}(n) -\alpha\sigma g_{-}(n-1)f_{+}(n)\bar{r}_{+}(n)
\end{eqnarray}
\begin{eqnarray}\label{eq-6.7}
	&& +\mathrm{i}\frac{\mathrm{d}}{\mathrm{d}\tau}\bar{q}_{-}(n) = -\beta q_{+}(n) - \alpha  q_{+}(n+1)[1+\sigma r_{-}(n)\bar{q}_{-}(n)] - \beta\sigma\nu(n)\bar{q}_{-}(n)\nonumber\\ 
	&& -\alpha\sigma \bar{q}_{-}(n)[\mu(n)-r_{+}(n)\bar{q}_{-}(n)] - \beta\sigma \bar{q}_{-}(n)r_{+}(n+1)[q_{-}(n)-\bar{q}_{-}(n)]\nonumber\\
	&& -\beta\sigma f_{-}(n)g_{+}(n+1)\bar{q}_{-}(n) -\alpha\sigma f_{+}(n+1)g_{-}(n)\bar{q}_{-}(n)
\end{eqnarray}
\begin{eqnarray}\label{eq-6.8}
	&& -\mathrm{i}\frac{\mathrm{d}}{\mathrm{d}\tau}\bar{r}_{-}(n) = -\alpha r_{+}(n) - \beta  r_{+}(n+1)[1+\sigma q_{-}(n)\bar{r}_{-}(n)] - \alpha\sigma\mu(n)\bar{r}_{-}(n)\nonumber\\ 
	&& -\beta\sigma \bar{r}_{-}(n)[\nu(n)-q_{+}(n)\bar{r}_{-}(n)] - \alpha\sigma \bar{r}_{-}(n)q_{+}(n+1)[r_{-}(n)-\bar{r}_{-}(n)]\nonumber\\
	&& -\alpha\sigma g_{-}(n)f_{+}(n+1)\bar{r}_{-}(n) -\beta\sigma g_{+}(n+1)f_{-}(n)\bar{r}_{-}(n)
\end{eqnarray}

\begin{eqnarray}\label{eq-6.9}
	&& +\mathrm{i}\sigma\frac{\mathrm{d}}{\mathrm{d}\tau}\ln[f_{+}(n)] = -\alpha \mu(n) + \beta q_{+}(n)\bar{r}_{-}(n)\nonumber\\ 
	&& + \beta q_{-}(n-1)r_{+}(n) - \alpha q_{+}(n)r_{-}(n-1) - \alpha q_{+}(n+1)[r_{-}(n)-\bar{r}_{-}(n)]\nonumber\\
	&& - \alpha f_{+}(n+1)g_{-}(n) - \alpha f_{+}(n)g_{-}(n-1) + \beta f_{-}(n-1)g_{+}(n)
\end{eqnarray}
\begin{eqnarray}\label{eq-6.10}
	&& -\mathrm{i}\sigma\frac{\mathrm{d}}{\mathrm{d}\tau}\ln[g_{+}(n)] = -\beta \nu(n) + \alpha r_{+}(n)\bar{q}_{-}(n)\nonumber\\ 
	&& + \alpha r_{-}(n-1)q_{+}(n) - \beta r_{+}(n)q_{-}(n-1) - \beta r_{+}(n+1)[q_{-}(n)-\bar{q}_{-}(n)]\nonumber\\
	&& - \beta g_{+}(n+1)f_{-}(n) - \beta g_{+}(n)f_{-}(n-1) + \alpha g_{-}(n-1)f_{+}(n)
\end{eqnarray}
\begin{eqnarray}\label{eq-6.11}
	&& +\mathrm{i}\sigma\frac{\mathrm{d}}{\mathrm{d}\tau}\ln[f_{-}(n)] = -\beta \nu(n) + \alpha q_{-}(n)\bar{r}_{+}(n)\nonumber\\ 
	&& + \alpha q_{+}(n+1)r_{-}(n) - \beta q_{-}(n)r_{+}(n+1) - \beta q_{-}(n-1)[r_{+}(n)-\bar{r}_{+}(n)]\nonumber\\
	&& - \beta f_{-}(n-1)g_{+}(n) - \beta f_{-}(n)g_{+}(n+1) + \alpha f_{+}(n+1)g_{-}(n)
\end{eqnarray}
\begin{eqnarray}\label{eq-6.12}
	&& -\mathrm{i}\sigma\frac{\mathrm{d}}{\mathrm{d}\tau}\ln[g_{-}(n)] = -\alpha \mu(n) + \beta r_{-}(n)\bar{q}_{+}(n)\nonumber\\ 
	&& + \beta r_{+}(n+1)q_{-}(n) - \alpha r_{-}(n)q_{+}(n+1) - \alpha r_{-}(n-1)[q_{+}(n)-\bar{q}_{+}(n)]\nonumber\\
	&& - \alpha g_{-}(n-1)f_{+}(n) - \alpha g_{-}(n)f_{+}(n+1) + \beta g_{+}(n+1)f_{-}(n)\,.
\end{eqnarray}
Here the concomitant field functions $\mu(n)$ and $\nu(n)$ are given by formulas
\begin{eqnarray}\label{eq-6.13}
	\mu(n) = q_{-}(n)\bar{r}_{+}(n)+\bar{q}_{-}(n)r_{+}(n)-\bar{q}_{-}(n)\bar{r}_{+}(n)
\end{eqnarray} 
\begin{eqnarray}\label{eq-6.14}
	\nu(n) = r_{-}(n)\bar{q}_{+}(n)+\bar{r}_{-}(n)q_{+}(n)-\bar{r}_{-}(n)\bar{q}_{+}(n)\,.
\end{eqnarray}
We would like to remind that the spatially independent parameters $\alpha$ and $\beta$ can be arbitrary functions of time, while the parameter $\sigma$ defined as $\sigma^{2}=1$ serves to distinguish two types of nonlinearities.

It is worth noticing that the twelve-component system (\ref{eq-6.1})--(\ref{eq-6.14}) can be treated as settled on two mutually coupled chains marked by indices $+$ and $-$. In this sense the lattice as a whole is proved to be a quasi-one-dimensional one. The  system's  spatial quasi-one-dimensionality and multicomponentness  could  be very prospective attributes in modeling transport  properties of long macromolecules both natural   and synthetic   origins  \cite{perepelytsya-EBJ-50-759,  satchanska-PM-16-1159,  chen-AM-32-2001893}.

The obtained dynamical system (\ref{eq-6.1})--(\ref{eq-6.14}) admits at least two types of  symmetries.

\subsection{Symmetry under the  complex conjugation}
Thus, the system's symmetry under the  complex conjugation is based on the   complex conjugate symmetries of  field functions
\begin{eqnarray}\label{eq-6.15}
	 r^{*}_{+}(n)=q_{+}(n)
\end{eqnarray}
\begin{eqnarray}\label{eq-6.16}
	 r^{*}_{-}(n)=q_{-}(n)
\end{eqnarray}
\begin{eqnarray}\label{eq-6.17}
	 \bar{r}^{*}_{+}(n)=\bar{q}_{+}(n)
\end{eqnarray}
\begin{eqnarray}\label{eq-6.18}
	 \bar{r}^{*}_{-}(n)=\bar{q}_{-}(n)
\end{eqnarray}
\begin{eqnarray}\label{eq-6.19}
	 g^{*}_{+}(n)=f_{+}(n) 
\end{eqnarray}
\begin{eqnarray}\label{eq-6.20}
	 g^{*}_{-}(n)=f_{-}(n)
\end{eqnarray}
\begin{eqnarray}\label{eq-6.21}
	\nu^{*}(n)=\mu(n)
\end{eqnarray}
that are valid provided  the  parameters $\alpha$,  $-\mathrm{i}$, $\beta$  
meet  the conditions of complex conjugation 
\begin{eqnarray}\label{eq-6.22}
	 \beta^{*}=\alpha 
\end{eqnarray}
\begin{eqnarray}\label{eq-6.23}
	(-\mathrm{i})^{*}=+\mathrm{i}\,. 
\end{eqnarray}
Due to their complexvalueness the coupling parameters $\alpha$ and $\beta$ are able to model the impact of  external uniform magnetic field in terms of Peierls phase factors \cite{peierls-ZPh-80-763,  feynman-FLP3-1963,   vakhnenko-PRE-77-026604}.

\subsection{Symmetry under the  space and time reversal}
The system's symmetry under the  space and time reversal   is more sophisticated and it is based on the dynamical  properties of following transformed field functions 
\begin{eqnarray}\label{eq-6.24}
	\mathrm{q}_{+}(n)\equiv\mathrm{q}_{+}(n|\tau)=r_{-}(-n|-\tau)
\end{eqnarray}
\begin{eqnarray}\label{eq-6.25}
	\mathrm{r}_{+}(n)\equiv\mathrm{r}_{+}(n|\tau)=q_{-}(-n|-\tau)
\end{eqnarray}
\begin{eqnarray}\label{eq-6.26}
	\mathrm{q}_{-}(n)\equiv\mathrm{q}_{-}(n|\tau)=r_{+}(-n|-\tau)
\end{eqnarray}
\begin{eqnarray}\label{eq-6.27}
	\mathrm{r}_{-}(n)\equiv\mathrm{r}_{-}(n|\tau)=q_{+}(-n|-\tau)
\end{eqnarray}
\begin{eqnarray}\label{eq-6.28}
	\bar{\mathrm{q}}_{+}(n)\equiv\bar{\mathrm{q}}_{+}(n|\tau)=\bar{r}_{-}(-n|-\tau)
\end{eqnarray}
\begin{eqnarray}\label{eq-6.29}
	\bar{\mathrm{r}}_{+}(n)\equiv\bar{\mathrm{r}}_{+}(n|\tau)=\bar{q}_{-}(-n|-\tau)
\end{eqnarray}
\begin{eqnarray}\label{eq-6.30}
	\bar{\mathrm{q}}_{-}(n)\equiv\bar{\mathrm{q}}_{-}(n|\tau)=\bar{r}_{+}(-n|-\tau)
\end{eqnarray}
\begin{eqnarray}\label{eq-6.31}
	\bar{\mathrm{r}}_{-}(n)\equiv\bar{\mathrm{r}}_{-}(n|\tau)=\bar{q}_{+}(-n|-\tau)
\end{eqnarray}
\begin{eqnarray}\label{eq-6.32}
	\bar{\mathrm{f}}_{+}(n)\equiv\bar{\mathrm{f}}_{+}(n|\tau)=g_{-}(-n|-\tau)
\end{eqnarray}
\begin{eqnarray}\label{eq-6.33}
	\bar{\mathrm{g}}_{+}(n)\equiv\bar{\mathrm{g}}_{+}(n|\tau)=f_{-}(-n|-\tau)
\end{eqnarray}
\begin{eqnarray}\label{eq-6.34}
	\bar{\mathrm{f}}_{-}(n)\equiv\bar{\mathrm{f}}_{-}(n|\tau)=g_{+}(-n|-\tau)
\end{eqnarray}
\begin{eqnarray}\label{eq-6.35}
	\bar{\mathrm{g}}_{-}(n)\equiv\bar{\mathrm{g}}_{-}(n|\tau)=f_{+}(-n|-\tau)
\end{eqnarray}
\begin{eqnarray}\label{eq-6.36}
\upmu (n)\equiv\upmu(n|\tau)=\mu(-n|-\tau)
\end{eqnarray}
\begin{eqnarray}\label{eq-6.37}
	\upnu(n)\equiv\upnu(n|\tau)=\nu(-n|-\tau)\,.
\end{eqnarray}
The simple comparison shows that the  transformed field functions (\ref{eq-6.24})--(\ref{eq-6.35}) are governed by the set of twelve semi-discrete nonlinear equations invariant to the twelve-component semi-discrete nonlinear integrable equations of our interest (\ref{eq-6.1})--(\ref{eq-6.12}). 

Here any additional requirements on parameters  $\alpha$,  $\mathrm{i}$, $\beta$  are seen to be unnecessary. In this sense  the space-time reversal symmetry of inspected  semi-discrete nonlinear integrable  system (\ref{eq-6.1})--(\ref{eq-6.12}) turns out to be  more general than  the usual parity-time ($\cal{PT}$) symmetry \cite{bender-RPP-70-947,  konotop-RMP-88-035002}.

\section{Six-component semi-discrete   nonlinear  integrable  system  and its symmetry under the  space  and time reversal}   
\label{sec7}
\setcounter{equation}{0}

The reduction formulas (\ref{eq-5.9})--(\ref{eq-5.16}) listed in {\it Subsection 5.2} as applied to the prototype set  of semi-discrete equations (\ref{eq-3.1})--(\ref{eq-3.16})
accompanied by the formulas referred in the first paragraph of Section 5 give rise to the following six-component semi-discrete  nonlinear  integrable  system   
\begin{eqnarray}\label{eq-7.1}
	&& \frac{\mathrm{d}}{\mathrm{d}\tau}w_{+}(n) = -\varkappa w_{-}(n) + \varkappa  w_{-}(n-1)[1+\sigma w_{+}(n)w_{+}(n)] - \varkappa\sigma\pi(n)w_{+}(n)\nonumber\\ 
	&& +\varkappa\sigma w_{+}(n+1)[\pi(n)-w_{-}(n)w_{+}(n)] - \varkappa\sigma h_{+}(n+1)h_{-}(n)[w_{+}(n)-\bar{w}_{+}(n)]\nonumber\\
	&& +\varkappa\sigma h_{+}(n)h_{-}(n-1)[w_{+}(n) - \bar{w}_{+}(n-1)]
\end{eqnarray}
\begin{eqnarray}\label{eq-7.2}
	&& \frac{\mathrm{d}}{\mathrm{d}\tau}w_{-}(n) = + \varkappa w_{+}(n) - \varkappa  w_{+}(n+1)[1+\sigma w_{-}(n)w_{-}(n)] + \varkappa\sigma\pi(n)w_{-}(n)\nonumber\\ 
	&& -\varkappa\sigma w_{-}(n-1)[\pi(n)-w_{+}(n)w_{-}(n)] + \varkappa\sigma h_{-}(n-1)h_{+}(n)[w_{-}(n)-\bar{w}_{-}(n)]\nonumber\\
	&& -\varkappa\sigma h_{-}(n)h_{+}(n+1)[w_{-}(n) -\bar{w}_{-}(n+1)]
\end{eqnarray}
  
\begin{eqnarray}\label{eq-7.3}
	&& \frac{\mathrm{d}}{\mathrm{d}\tau}\bar{w}_{+}(n) = -\varkappa w_{-}(n)[1+\sigma \bar{w}_{+}(n)\bar{w}_{+}(n)] + \varkappa  w_{-}(n-1)[1+\sigma \bar{w}_{+}(n)\bar{w}_{+}(n)]
\end{eqnarray}
\begin{eqnarray}\label{eq-7.4}
	&& \frac{\mathrm{d}}{\mathrm{d}\tau}\bar{w}_{-}(n) = +\varkappa w_{+}(n)[1+\sigma \bar{w}_{-}(n)\bar{w}_{-}(n)] - \varkappa  w_{+}(n+1)[1+\sigma \bar{w}_{-}(n)\bar{w}_{-}(n)] 
\end{eqnarray}

\begin{eqnarray}\label{eq-7.5}
	&& \sigma\frac{\mathrm{d}}{\mathrm{d}\tau}\ln[h_{+}(n)] = -\varkappa \pi(n) - \varkappa w_{+}(n)\bar{w}_{-}(n)  - 2\varkappa w_{-}(n-1)w_{+}(n)            \nonumber\\ 
	&&  - \varkappa w_{+}(n+1)[w_{-}(n)-\bar{w}_{-}(n)]
	- \varkappa h_{+}(n+1)h_{-}(n)  - 2\varkappa h_{-}(n-1)h_{+}(n)
\end{eqnarray}
\begin{eqnarray}\label{eq-7.6}
	&& \sigma\frac{\mathrm{d}}{\mathrm{d}\tau}\ln[h_{-}(n)] = +\varkappa \pi(n) + \varkappa w_{-}(n)\bar{w}_{+}(n) + 2\varkappa w_{+}(n+1)w_{-}(n) \nonumber\\ 
	&&  + \varkappa w_{-}(n-1)[w_{+}(n)-\bar{w}_{+}(n)]
	 + \varkappa h_{-}(n-1)h_{+}(n)  + 2\varkappa h_{+}(n+1)h_{-}(n)\,.
\end{eqnarray}
Here the concomitant field function $\pi(n)$ is given by formula
\begin{eqnarray}\label{eq-7.7}
	\pi(n)=w_{-}(n)\bar{w}_{+}(n)+\bar{w}_{-}(n)w_{+}(n)-\bar{w}_{-}(n)\bar{w}_{+}(n)\,.
\end{eqnarray}
The system's symmetry under the  space and time reversal    is based on the dynamical  properties of following transformed field functions 
\begin{eqnarray}\label{eq-7.8}
	\mathrm{w}_{+}(n)\equiv\mathrm{w}_{+}(n|\tau)=w_{-}(-n|-\tau)
\end{eqnarray}
\begin{eqnarray}\label{eq-7.25}
	\mathrm{w}_{-}(n)\equiv\mathrm{w}_{-}(n|\tau)=w_{+}(-n|-\tau)
\end{eqnarray}
\begin{eqnarray}\label{eq-7.26}
	\bar{\mathrm{w}}_{+}(n)\equiv\bar{\mathrm{w}}_{+}(n|\tau)=\bar{w}_{-}(-n|-\tau)
\end{eqnarray}
\begin{eqnarray}\label{eq-7.27}
	\bar{\mathrm{w}}_{-}(n)\equiv\bar{\mathrm{w}}_{-}(n|\tau)=\bar{w}_{+}(-n|-\tau)
\end{eqnarray}
\begin{eqnarray}\label{eq-7.28}
	\bar{\mathrm{h}}_{+}(n)\equiv\bar{\mathrm{h}}_{+}(n|\tau)=h_{-}(-n|-\tau)
\end{eqnarray}
\begin{eqnarray}\label{eq-7.29}
	\bar{\mathrm{h}}_{-}(n)\equiv\bar{\mathrm{h}}_{-}(n|\tau)=h_{+}(-n|-\tau)
\end{eqnarray}
\begin{eqnarray}\label{eq-7.30}
	\uppi (n)\equiv\uppi(n|\tau)=\pi(-n|-\tau)\,.
\end{eqnarray}
The simple comparison shows that the  transformed field functions (\ref{eq-7.8})--(\ref{eq-7.29}) are governed by the set of six semi-discrete nonlinear equations invariant to the  six-component semi-discrete nonlinear integrable equations of our  interest (\ref{eq-7.1})--(\ref{eq-7.6}).

\section{Discussion}   
\label{sec8}
\setcounter{equation}{0}

The semi-discrete nonlinear integrable systems (\ref{eq-6.1})--(\ref{eq-6.12}) and (\ref{eq-7.1})--(\ref{eq-7.6})  presented  in Section 6 and Section 7 are proved  to be essentially multicomponent ones inasmuch either of them cannot be split into several physically uncoupled subsystems. 

Thus, the twelve-component semi-discrete integrable system  (\ref{eq-6.1})--(\ref{eq-6.12}) is composed of six subsystems coupled by linear and nonlinear types of interactions. These subsystems are formalized by the six pairs of field functions. 
The set of plausible   pairs are as follows 
$q_{+}(n)\leftrightarrow r_{+}(n)$,~  $\bar{q}_{+}(n)\leftrightarrow \bar{r}_{+}(n)$,~  $f_{+}(n)\leftrightarrow g_{+}(n)$,~ $q_{-}(n)\leftrightarrow r_{-}(n)$,~  $\bar{q}_{-}(n)\leftrightarrow \bar{r}_{-}(n)$,~  $f_{-}(n)\leftrightarrow g_{-}(n)$. Here the symbol~ $\leftrightarrow$~  is inserted  to point out on  a suppositional canonical relationship between the field functions of a particular pair. The problems of adequate Hamiltonian treatment and reliable Poisson structure characterizing the twelve-component semi-discrete nonlinear integrable system (\ref{eq-6.1})--(\ref{eq-6.12})
appear to be very complicated and presently they are opened for the future investigations. The general principles of  Hamiltonian  and Poisson approaches as applied  to the multicomponent semi-discrete nonlinear integrable systems have been    approbated  in our previous papers \cite{vakhnenko-JNMP-24-250, vakhnenko-JMP-56-033505, vakhnenko-JMP-59-053504, vakhnenko-LMP-108-1807, vakhnenko-JMP-57-113504}.  The difficult aspects in establishing the Hamiltonian  and  Poisson structures for the systems of present interest (\ref{eq-6.1})--(\ref{eq-6.12}) and  (\ref{eq-7.1})--(\ref{eq-7.6}) are summarized in Section A. 

Though the  Hamiltonian treatment of  twelve-component semi-discrete integrable system (\ref{eq-6.1})--(\ref{eq-6.12}) is waiting for its rigorous formulation the analysis of universal local conservation law (\ref{eq-4.3}) provides us with certain fruitful information about plausible physical sense of  involved subsystems at least in the case of system's complex conjugate symmetry. 

To be precise, we should inspect the expressions for the   local conserved density $\rho(n)$ (\ref{eq-4.8})  and the local current $J(n)$ having been adapted to the needs of reduced semi-discrete nonlinear integrable system (\ref{eq-6.1})--(\ref{eq-6.12})  under the premise of complex conjugation symmetry  (\ref{eq-6.15})--(\ref{eq-6.23}).  The announced adapted quantities as well as the respective local conservation law read as follows 
\begin{eqnarray}\label{eq-8.1}
	\rho(n) = \ln[1+\sigma\bar{q}_{+}(n)\bar{r}_{+}(n)] +          \ln[1+\sigma\bar{q}_{-}(n)\bar{r}_{-}(n)] + \ln[f_{+}(n)g_{+}(n)f_{-}(n)g_{-}(n)]
\end{eqnarray} 
\begin{eqnarray}\label{eq-8.2}
	J(n)&=&\mathrm{i}\alpha\sigma q_{+}(n)r_{-}(n-1)+\mathrm{i}\alpha\sigma f_{+}(n)g_{-}(n-1)\nonumber\\
        &-&\mathrm{i}\beta\sigma r_{+}(n)q_{-}(n-1) - \mathrm{i}\beta\sigma g_{+}(n)f_{-}(n-1)
\end{eqnarray}
\begin{eqnarray}\label{eq-8.3}
	\frac{\mathrm{d}}{\mathrm{d}\tau}\rho(n) = J(n) - J(n+1)\,.
\end{eqnarray}  
Here the partial local densities  
\begin{eqnarray}\label{eq-8.4}
	\bar{\rho}_{+}(n) = \ln[1+\sigma\bar{q}_{+}(n)\bar{r}_{+}(n)] 
\end{eqnarray}
and
\begin{eqnarray}\label{eq-8.5}
	\bar{\rho}_{-}(n) = \ln[1+\sigma\bar{q}_{-}(n)\bar{r}_{-}(n)] 
\end{eqnarray}
are essentially separate characteristics related to subsystems 
    $\bar{q}_{+}(n)\leftrightarrow \bar{r}_{+}(n)$  and  $\bar{q}_{-}(n)\leftrightarrow \bar{r}_{-}(n)$, respectively. 
On the other hand, the net local density  
\begin{eqnarray}\label{eq-8.6}
	\rho_{\pm}(n) = \ln[f_{+}(n)g_{+}(n)f_{-}(n)g_{-}(n)]
\end{eqnarray}
is  related to  two  subsystems  $f_{+}(n)\leftrightarrow g_{+}(n)$ and  $f_{-}(n)  \leftrightarrow g_{-}(n)$ combined. 

In the case of   attractive nonlinearity  $\sigma=+1$ the first two local densities (\ref{eq-8.4}) and (\ref{eq-8.5}) are seen to be the real-valued nonnegative quantities treatable  as the local densities  of positive charges   associated with the respective  fields  $\bar{q}_{+}(n)\leftrightarrow \bar{r}_{+}(n)$  and  $\bar{q}_{-}(n)\leftrightarrow \bar{r}_{-}(n)$. In contrast, the third   real-valued quantity (\ref{eq-8.6}) can acquire either positive or negative magnitude. This quantity can be treated as the  net local density of charge related to  two  subsystems described by two pairs of fields $f_{+}(n)\leftrightarrow g_{+}(n)$ and  $f_{-}(n)  \leftrightarrow g_{-}(n)$. As to the total charge
\begin{eqnarray}\label{eq-8.7}
	Q = \sum_{m=-\infty}^\infty [\,\bar{\rho}_{+}(m) + \bar{\rho}_{-}(m) + \rho_{\pm}(m)\,]
\end{eqnarray}
accumulated in the whole twelve-component system (\ref{eq-6.1})--(\ref{eq-6.12}),  it is seen to be conserved provided  the local current  $J(n)$ is the same on both spatial infinities.

The case of repulsive nonlinearity $\sigma=-1$  turns out to be more complicated, inasmuch as now the clear physical treatment of densities $\bar{\rho}_{+}(n)$ and $\bar{\rho}_{-}(n)$ as the nonpositive charge densities  is enabled only under the  strict limitations 
\begin{eqnarray}\label{eq-8.8}
	0 \leq \bar{q}_{+}(n)\bar{r}_{+}(n) < 1
\end{eqnarray}
\begin{eqnarray}\label{eq-8.9}
	0 \leq \bar{q}_{-}(n)\bar{r}_{-}(n) < 1\,.
\end{eqnarray}
It is presently unknown whether or not these restrictions are globally achievable under certain type of special boundary conditions  similar to those suitable  for  the usual  semi-discrete nonlinear Schr\"{o}dinger system with the repulsive nonlinearity \cite{chubykalo-PLA-169-359,  vekslerchik-IP-8-889, konotop-JPA-25-4037}. In this situation the treatment of densities $\bar{\rho}_{+}(n)$ and $\bar{\rho}_{-}(n)$ as the nonpositive charge densities appears to be very conditional.   Nevertheless, the  total charge (\ref{eq-8.7}) 
accumulated in the whole twelve-component system (\ref{eq-6.1})--(\ref{eq-6.12}) is obliged  to be conserved even despite of its rather conditional physical treatment.  

Meanwhile, the   mathematical structure  of expression (\ref{eq-8.2}) for the total local current $J(n)$ indicates that only four of six  subsystems actually participate  in the charge transportation. They are described by four pairs of fields $q_{+}(n)\leftrightarrow r_{+}(n)$,~  $f_{+}(n)\leftrightarrow g_{+}(n)$,~ $q_{-}(n)\leftrightarrow r_{-}(n)$,~    $f_{-}(n)\leftrightarrow g_{-}(n)$. 

The six-component nonlinear integrable system   is formalized by six linearly and nonlinearly coupled semi-discrete equations (\ref{eq-7.1})--(\ref{eq-7.6}) for six field functions $w_{+}(n)$, $\bar{w}_{+}(n)$, $h_{+}(n)$, $w_{-}(n)$, $\bar{w}_{-}(n)$, $h_{-}(n)$. However in the case of  six-component system we \textit{a priori} unable to claim for the plausible pairs of presumably canonical field functions. 

The fields marked by plus subscript can be treated as settled on plus labeled chain, while the fields marked by minus subscript can be treated as settled on minus labeled chain of quasi-one-dimensional regular lattice regardless of whether the system under consideration is a twelve-component  or a six-component one.  In  this regard  the suggested twelve-component   semi-discrete nonlinear integrable system (\ref{eq-6.1})--(\ref{eq-6.12}) is proved to be  very prospective tool for modeling the physical properties of  multicomponent essentially quasi-one-dimensional  latticed objects under the action of  external uniform magnetic field and external parametric drive encodeable  in its  coupling parameters.

\section{Conclusion}   
\label{sec9}
\setcounter{equation}{0}

In our research we proposed two novel multicomponent semi-discrete nonlinear integrable systems prospective   for   modeling the transport phenomena in regular quasi-one-dimensional structures of both natural and synthetic origins. 

We expect the comprehensive investigation of these  semi-discrete nonlinear integrable systems will be interesting both from the physical and mathematical standpoints. 
Presently, the most evident open  problems  are  (1) to  construct   the rigorous analytical solutions,  and   (2) to disclose  the Hamiltonian and Poisson structures typifying  the suggested  semi-discrete nonlinear integrable systems. 

In our opinion, the most straightforward way to obtain  the explicit analytical solutions to  multicomponent semi-discrete nonlinear integrable systems is based upon the Darboux--B\"{a}cklund transformation technique \cite{vakhnenko-JNMP-24-250, vakhnenko-EPJP-137-1176, vakhnenko-JMP-56-033505, vakhnenko-JMP-59-053504, vakhnenko-PRSA-477-20210562, vakhnenko-PRE-108-024223}. 

The problem to establish the  Hamiltonian and Poisson structures of proposed integrable systems turn out to be immensely more complicated as compared with the analogous rather nontrivial  problems successfully  solved in our previous works \cite{vakhnenko-JNMP-24-250, vakhnenko-JMP-56-033505, vakhnenko-JMP-59-053504,  vakhnenko-LMP-108-1807, vakhnenko-JMP-57-113504}.
Some  aspects of  our preliminary approach to these problems  are  reported in  Section A.


\appendix 

\section{Preliminaries to Hamiltonian treatment}

Having observed  that all intersite interactions in the   twelve-component semi-discrete nonlinear integrable system (\ref{eq-6.1})--(\ref{eq-6.12}) are of nearest-neighbouring   
type it is reasonable to construct the system's Hamiltonian function relying upon  the local conserved densities (\ref{eq-4.29})--(\ref{eq-4.32}) characterized by the same type 
of couplings between the involved fields. As a result we come to the following Hamiltonian function
\begin{eqnarray}\label{eq-A.1}
 H = &-&\alpha \sum_{m=-\infty}^\infty [\,q_{-}(m)\bar{r}_{+}(m)+\bar{q}_{-}(m)r_{+}(m)-\bar{q}_{-}(m)\bar{r}_{+}(m)\,]\nonumber\\
 &-&\alpha \sum_{m=-\infty}^\infty [\,q_{+}(m)r_{-}(m-1)+q_{-}(m)r_{+}(m)+f_{+}(m)g_{-}(m-1)\,]\nonumber\\
 &-&\beta \sum_{m=-\infty}^\infty [\,r_{-}(m)\bar{q}_{+}(m)+\bar{r}_{-}(m)q_{+}(m)-\bar{r}_{-}(m)\bar{q}_{+}(m)\,]\nonumber\\
 &-&\beta \sum_{m=-\infty}^\infty [\,r_{+}(m)q_{-}(m-1)+r_{-}(m)q_{+}(m)+g_{+}(m)f_{-}(m-1)\,]\,.
\end{eqnarray}

According to the general rules \cite{maschke-JFI-329-923, torres del castillo-RMF-50-608},  the Hamiltonian  dynamic equations of motion for the system under study (\ref{eq-6.1})--(\ref{eq-6.12})  must be  sought  in the form
\begin{eqnarray}\label{eq-A.2}
	\frac{\mathrm{d}}{\mathrm{d}\tau}y_\lambda(n) = \sum_{\varkappa=1}^{12}  \sum_{m=-\infty}^\infty J_{\lambda\varkappa}(n|m)
	\frac{\partial H}{\partial y_{\varkappa}(m)}\qquad\qquad\qquad\qquad (\lambda = 1, 2, 3, ..., 12)\,,
\end{eqnarray}
where    $J_{\lambda\varkappa}(n|m)$ are the elements  of skew-symmetric $J_{\varkappa\lambda}(m|n)=-J_{\lambda\varkappa}(n|m)$ symplectic matrix. These elements $J_{\lambda\varkappa}(n|m)$ are obliged to define the  Poisson bracket 
\begin{eqnarray}\label{eq-A.3}
	\{F,G\} = -\sum_{n=-\infty}^\infty \,\sum_{\lambda=1}^{12} \,\sum_{\varkappa=1}^{12} \,\sum_{m=-\infty}^\infty \frac{\partial F}{\partial y_{\lambda}(n)} J_{\lambda\varkappa}(n|m) \frac{\partial G}{\partial y_{\varkappa}(m)}
\end{eqnarray}
subjected to the Jacobi identity
\begin{eqnarray}\label{eq-A.4}
	\{E,\{F,G \} \} +\{F,\{G,E \} \} +\{G,\{E,F \} \}=0\,.
\end{eqnarray}
In so doing, the set of Hamiltonian equations (\ref{eq-A.2}) acquires the form 
\begin{eqnarray}\label{eq-A.5}
	\frac{\mathrm{d}}{\mathrm{d}\tau}y_\lambda(n) = \{H,\,y_{\lambda}(n)\}\qquad\qquad\qquad\qquad (\lambda = 1, 2, 3, ..., 12)
\end{eqnarray}
substantiated by the set of sixty six fundamental Poisson brackets  
\begin{eqnarray}\label{eq-A.6}
\{ y_{\lambda}(n),\, y_{\varkappa}(m)\} = -  J_{\lambda\varkappa}(n|m)\,. 
\end{eqnarray}

We tried to apply the above described procedure (\ref{eq-A.2})--(\ref{eq-A.6}) to our twelve-component integrable system (\ref{eq-6.1})--(\ref{eq-6.12}) relying on the adopted Hamiltonian  function (\ref{eq-A.1}) and introducing the  universal notations 
\begin{eqnarray}\label{eq-A.7}
	y_{1}(n) = q_{-}(n)\qquad\qquad\qquad\qquad\qquad\qquad y_{7}(n) = r_{-}(n)
\end{eqnarray}
\begin{eqnarray}\label{eq-A.8}
	y_{2}(n) = q_{+}(n)\qquad\qquad\qquad\qquad\qquad\qquad y_{8}(n) = r_{+}(n) 
\end{eqnarray}
\begin{eqnarray}\label{eq-A.9}
	y_{3}(n) = \bar{q}_{-}(n)\qquad\qquad\qquad\qquad\qquad\qquad y_{9}(n) = \bar{r}_{-}(n)
\end{eqnarray}
\begin{eqnarray}\label{eq-A.10}
	y_{4}(n) = \bar{q}_{+}(n)\qquad\qquad\qquad\qquad\qquad\qquad y_{10}(n) = \bar{r}_{+}(n)
\end{eqnarray}
\begin{eqnarray}\label{eq-A.11}
	y_{5}(n) = f_{-}(n)\qquad\qquad\qquad\qquad\qquad\qquad y_{11}(n) = g_{-}(n)
\end{eqnarray}
\begin{eqnarray}\label{eq-A.12}
	y_{6}(n) = f_{+}(n)\qquad\qquad\qquad\qquad\qquad\qquad y_{12}(n) = g_{+}(n)
\end{eqnarray}
for the system's field functions. However, we have not managed  to isolate extremally huge number of  candidates on the elements $J_{\lambda\varkappa}(n|m)$ of symplectic matrix to say nothing about their verification via the Jacobi identity (\ref{eq-A.4}) or its more simple equivalents \cite {maschke-JFI-329-923, torres del castillo-RMF-50-608, vakhnenko-JMP-59-053504, vakhnenko-LMP-108-1807}.

The main obstacle  in achieving the positive result is the generic spatial nonlocality  
of inspected symplectic matrix $J_{\lambda\varkappa}(n|m)$  pronouncedly  contrasting with the crucial simplification 
 \begin{eqnarray}\label{eq-A.13}
 	J_{\lambda\varkappa}(n|m) = J_{\lambda\varkappa}(n|n)\delta_{nm}
\end{eqnarray}
typical of earlier studied systems  \cite{vakhnenko-JNMP-24-250, vakhnenko-JMP-56-033505, vakhnenko-JMP-59-053504,  vakhnenko-LMP-108-1807, vakhnenko-JMP-57-113504}.

\subsection*{Acknowledgements}
\addcontentsline{toc}{section}{Acknowledgements}

Oleksiy O. Vakhnenko acknowledges  support from the National Academy of Sciences of Ukraine within the Project No~0122U000887.  
Vyacheslav O. Vakhnenko acknowledges  support from the National Academy of Sciences of Ukraine within the Project No~0123U100182.
Oleksiy O. Vakhnenko also acknowledges  support from the Simons Foundation (USA) under the Grant SFI-PD-Ukraine-00014573. 
We are grateful to  Reviewers for the constructive criticism directed to improve the quality of presented results. 



\addcontentsline{toc}{section}{References}

\pdfbookmark[1]{References}{ref}
\LastPageEnding

\end{document}